# Design of Quantum Key Agreement Protocols with Strong Fairness Property

Kun-Fei Yu[1], Chun-Wei Yang[2], Tzonelih Hwang[*,1], Chuan-Ming Li[3] and Jun Gu[1]

[1]*Department of Computer Science and Information Engineering, National Cheng Kung University, No. 1, University Rd., Tainan City, 70101, Taiwan, R.O.C.*

[2]*Smart Network System Institute，Institute for Information Industry，Taipei City，Taiwan, ROC*

[3]*Department of Information Management, Shu-Zen Junior College of Medicine and Management, No. 452, Huanqiu Rd., Luzhu Dist., Kaohsiung 821, Taiwan, ROC*

[*]**Responsible for correspondence:**

Tzonelih Hwang

Distinguished Professor

Department of Computer Science and Information Engineering,

National Cheng Kung University,

No. 1, University Rd.,

Tainan City, 70101, Taiwan, R.O.C.

Email: hwangtl@ismail.csie.ncku.edu.tw

TEL: +886-6-2757575 ext. 62524

[*]corresponding author

# Abstract

This study distinguishes the weak fairness property from the strong fairness property which is necessary in the quantum key agreement (QKA) and shows that most of the existing QKAs cannot achieve the strong fairness property with a key manipulation problem. To solve this problem, a model which describes the way to design a QKA with the strong fairness property is proposed. Based on the model, an example QKA is presented. Security analyses show that the proposed QKA is effective to resist not only outsider's eavesdropping attack but also insider's key manipulation.

# 1 Introduction

Key distribution (KD) is one of the most useful cryptographic research topics where the involved participants can securely share a secret key which can be used for further secure communications. Most researches of KD focus on how to achieve the **security property** (share a secure key among the involved participants). However, in the KD, the content of the shared secret key is usually determined by one of participants and then distributed to the others. Different from the KD, the key agreement (KA) protocols try to let each participant to contribute his/her influence to the final shared key. The first secure and fair KA protocol is proposed by Diffe and Hellman in 1976 [1]. Subsequently, several researches about KA have been proposed and the research results have been widely used, e.g. in Secure Sockets Layer (SSL) [2] and in Transport Layer Security (TLS) [2]. In order to standardize the KA, the International Organization for Standardization proposed a rigorous definition of KA in ISO/IEC DIS 11770-3 as follows [3].

KA is a process of establishing a shared secret key between entities in such a way that neither of them can predetermine the value of that key. (NOTE: By predetermine it is meant that neither entity A nor entity B can, in a computationally efficient way, choose a smaller key space and force the computed key in the protocol to fall into that key space.)

That is, in KA, the final shared key must be determined by all the involved participants. Any proper subset of the participants cannot determine or manipulate the final shared key alone. Hence, the KA protocols should not only achieve the security property but also achieve the fairness property. To achieve the security property, these proposed KA protocols are designed based on computation complexity. However, with the development of quantum computer which has excellent computational power, these proposed KA protocols face serious secure challenges [4, 5]. Hence, several researches of quantum key agreement protocol (QKA) where the security is guaranteed by the laws of quantum mechanics have been proposed.

In 2004, Zhou *et al.* [6] employed quantum teleportation technique and maximally entangled states to propose the first QKA. They claimed their protocol can ensure the participants to share a fair and secure key. However, Tsai *et al.* [7] commented that Zhou *et al.*'s protocol is susceptible to insider attacks. That is, a participant has the ability to fully control the secret key without being detected by the other. Therefore, the fairness property are not satisfied in Zhou *et al*'s scheme. In 2010, Chong *et al.* [8] proposed a QKA based on the BB84 protocol [9]. They pointed out that a malicious participant, Bob, can control one bit of the key in their protocol. In 2013, several multi-party QKAs [10-12] were presented, which allow numerous participants to take part in their communication. Subsequently, Huang *et al.* [13] proposed a new QKA using Einstein-Podolsky-Rosen (EPR) pairs and single-particle measurements. In 2014, Huang et al. [14] showed Sun et al.'s protocol cannot ensure each participant to share a secure and fair key and pointed out that it is necessary to concern the fairness property in the classical postprocessing process where each participant corrects the raw key to obtain the final key. Without designing a fair classical postprocessing process, the final shared key cannot be directly used in real life. Subsequently, Huang et al. [13] proposed a method to share EPR pairs over two collective-noise channels and Yuan et al. [15] designed an identity authentication method. Both of the methods can improve the fairness of QKAs in the classical postprocessing process. Obviously, all the existing related researches of QKA [8, 10-14, 16-37] try to achieve both of the security property and the fairness property.

To further explain the concept of the security and the fairness property clearly, it indeed requires to define the term 'security' and the term 'fairness' carefully. Moreover, according to the definition in ISO/IEC DIS 11770-3, the existing definition of 'fairness' in QKA is not rigorous. This not rigorous definition makes most of the exiting 'fairness' QKA protocols cannot achieve the real fairness property. To solve these problems, this paper proposes the **security property**, the **weak fairness property** (existing definition of 'fairness' in QKAs) and the **strong fairness property** (new definition of 'fairness' based on ISO/IEC DIS 11770-3) and defines respectively as follows:

◆ *Security property*: The outside eavesdropper cannot get any useful information of the final shared key without being detected and the protocol must ensure the involved participants to share an identical key.

◆ *Weak fairness property*: Each participant contributes his/her influence to the final shared key.

◆ *Strong fairness property*: Each participant does not have a single bit of advantage over the others to determine the final shared key. That is, none of the participant can manipulate even one bit of the final key.

According to these definitions, QKAs in [8, 10-14, 16-22, 24-30, 32-35, 38]

cannot achieve the strong fairness property. That is, in the protocols, a legitimate but malicious participant has the chance to derive the final shared secret key first during the public discussion process. If the malicious participant does not like the negotiated shared secret key to be the final session key, he/she can deliberately abort the protocol and then impute the error to an outside eavesdropper without being detected. By this way, these QKAs are vulnerable to a key manipulation problem, which certainly fails the strong fairness definition in a QKA. The purpose of this paper is to present a solution model to solve this problem. By using the proposed solution model, the designed QKAs can achieve strong fairness property. For convenience, here we just take Huang *et al.*'s [13] QKA as an example to demonstrate the key manipulation problem in detail.

The rest of this paper is organized as follows. Section 2 reviews Huang *et al.*'s QKA and subsequently describes the key manipulation problem. Section 3 introduces the proposed model, and presents a two-party QKA as well. Besides, security and fairness analyses of the proposed QKA are provided. Finally, a conclusion is given in Section 4.

## 2 Discussion of fairness property in existing QKAs

According to our research, several existing QKAs [8, 10-14, 16-22, 24-30, 32-35, 38] have a key manipulation problem which makes these QKAs cannot achieve the strong fairness property. Here, as an example, we use Huang *et* al.'s QKA protocol [13] to demonstrate it.

### 2.1 Review of Huang *et al.*'s QKA

Assume that there are two participants, Alice and Bob, who want to share a fair $n$-bit secret key by using EPR pairs and single-particle measurements. The two polarization bases, Z-basis ($\{|0\rangle, |1\rangle\}$) and X-basis ($\{|+\rangle, |-\rangle\}$), are used as the initial states of single particles, where $|+\rangle = \frac{1}{\sqrt{2}}(|0\rangle + |1\rangle)$ and $|-\rangle = \frac{1}{\sqrt{2}}(|0\rangle - |1\rangle)$. The protocol proceeds as follow:

**Step 1** Alice generates a sequence of $n$ Bell states, $S = \{s_1, s_2, \ldots, s_n\}$, in

$$|\Phi^+\rangle = \frac{1}{\sqrt{2}}(|00\rangle + |11\rangle) = \frac{1}{\sqrt{2}}(|++\rangle + |--\rangle)$$

where $s_i = \{q_A^i, q_B^i\}$, for $i = 1, 2, \ldots, n$. She divides $S$ into two sequences $S_A = \{q_A^i\}$ and $S_B = \{q_B^i\}$, for $i = 1, 2, \ldots, n$.

**Step 2** To prevent the transmitted particles from eavesdropping attacks, Alice prepares a sufficient number of decoy particles randomly in one of the four polarization states $\{|0\rangle, |1\rangle, |+\rangle, |-\rangle\}$. Then she randomly inserts these decoy particles into the sequence $S_B$ (denote the new sequence as $S_B^*$) and sends the sequence $S_B^*$ to Bob.

**Step 3** After confirming that Bob has received the sequence $S_B^*$, Alice announces the positions and the initial states of the decoy particles to Bob. Based on Alice's information, Bob picks the decoy particles out from $S_B^*$ and measures these decoy particles in the corresponding bases of the initial states. Then, Bob compares his measurement results with the initial states. If the error rate is higher than a pre-defined threshold, Alice and Bob abort the protocol and start a new one. Otherwise, they continue the next step.

**Step 4** Bob determines a string $C \in \{0,1\}^n$ and sends it to Alice. Based on the string $C$, they perform single-particle measurements on each particle. More precisely, if the $i$-th bit of $C$ is "0", Alice and Bob perform Z-basis measurements on the particle $q_A^i$ and $q_B^i$, respectively. Otherwise, they perform X-basis measurements. If the measurement result is $|0\rangle$ or $|+\rangle$ ($|1\rangle$ or $|-\rangle$), Alice and Bob can derive a value "0" ("1") as their shared key bit.

It appears that the shared secret key cannot be determined alone by any one of the participants because both participants measure all the particles with identical bases and subsequently get the same measurement results. However, the next section will show that in fact, one of the participants has the advantage to manipulate the final key.

## 2.2 Key manipulation problem

Suppose that there is a malicious participant, Bob, who attempts to control one-bit of the shared secret key. In Step 3 of Huang *et al.*'s protocol, Bob separates the decoy particles from the sequence $S_B^*$ after knowing the positions of the decoy particles in Step3. Bob preserves these decoy particles in quantum storage and instead directly measures the remaining particles (i.e., $S_B$) based on the string *C* that he determined in Step 4. As a result, Bob can derive the shared secret key earlier than Alice. Suppose for some reason, Bob prefers the first bit of the negotiated key to be "0", but in fact it turns out to be a "1". Then he can deliberately notify Alice that the comparison result in the public discussion is negative. Accordingly, the protocol will be aborted and they will start a new one until Bob gets the desired key bit to be a "1". In this way, even though the unpredictability of the entangled Bell states is applied for the key negotiation, Bob can still completely determine one bit of the key or even the whole key bits.

It would be futile even if the eavesdropper detection in the public discussion is modified to be performed by Alice instead of by Bob as in the original design. To explain this, let us modify the public discussion (Step 3) as the following:

**Step 3\*** After confirming that Bob has already received $S_B^*$, Alice announces the positions and the bases of the decoy particles to Bob. Based on Alice's announcement, Bob measures these decoy particles and then sends his measurement results back to Alice. Alice compares Bob's reports with the initial states that she prepared. If the error rate exceeds the pre-defined threshold, then the protocol will be aborted. Otherwise, they continue to the next step.

In this case, Bob still can perform the similar manipulation plot as below: upon receiving Alice's announcement, Bob preserves the decoy photons in the quantum storage. Then he directly measures the sequence $S_B$ based on the string $C$. Thus, Bob can first derive the shared secret key. In this regard, if Bob prefers this shared key, he will follow the protocol. Otherwise, Bob sends fake measurement results to Alice for eavesdropping check and definitely the comparison cannot be passed. As a result, the protocol will be started again until Bob gets a desired key.

By this way, the malicious participant, Bob, can easily destroy the fairness of the protocol by publishing a fake message during the public discussion. However, Alice has no way of detecting this fraudulence. Obviously, Bob can easily manipulate the final key of a key-agreement key and attribute all the fault to an eavesdropper without being detected. Similarly, the existing QKAs [8, 10-13, 39, 40] also have the similar problem. For simplicity, we omit the discussion here.

## 3  Design of QKAs with strong fairness

This section first proposes a solution model to avoid the key manipulation problem which we have demonstrated in Section 2.2. Then, we present an example of QKA based on the proposed model in Section 3.2. Security analyses of the proposed QKA are also provided in Section 3.3. Moreover, we discuss how the proposed scheme accomplishes the strong fairness property.

## 3.1 Solution model for QKA with strong fairness

Based on the discussion in Section 2, we summarize that in order to achieve the strong fairness property, a QKA has to satisfy the following requirements in the designing phase: (1) No participant is able to derive the final secret key before the public discussion is finished (2) Any manipulation on the shared key will be detected by the other participants.

Based on the above requirements, the proposed solution model for a fair QKA is divided into three stages as follows (also shown in Figure 1):

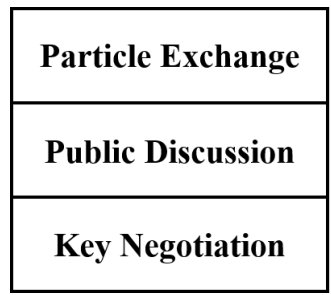

Figure 1. Solution model for a QKA with fairness

**Particle Exchange Stage:** the participants send out the sequence of particles or receive the corresponding sequence in this stage. It is required that every participant has to complete the particles transmission here.

**Public Discussion Stage:** the participants check the security of channel in this stage. This stage makes sure that the transmitted sequences are free from the outside eavesdropping attacks.

**Key Negotiation Stage:** the participants help each other to construct or derive the final secret key (here transmission of additional classical messages is needed). Note that the shared secret key can only be derived by the participants in this stage.

The main idea of the model is that the shared secret key cannot be derived until the participants have confirmed that the channel is free from the eavesdropping attacks. In other words, a QKA with strong fairness has to prevent any participant from obtaining the shared secret key before completing the public discussion.

## 3.2 New QKA based on the proposed solation model

Based on the above solution model, we design a two-party QKA as an example. Suppose that there are two parties, Alice and Bob, who would like to negotiate a shared secret key. Here, a hash function $H$ [41, 42] is used for eavesdropping check and for verifying the integrity of a secret message. Based on the property of a hash function, one-bit error in the input will cause significant changes in the output and can be detected. This property is very useful in checking message integrity if the quantum channel is reliable or ideal. Moreover, Alice and Bob agree on the following encoding rules: a binary value "0" is encoded as one of the two polarization states $\{|0\rangle, |+\rangle\}$ and a value "1" is encoded as $\{|1\rangle, |-\rangle\}$, and vice versa. Now, let us introduce the process of the new QKA in detail as follows (shown in Figure 2):

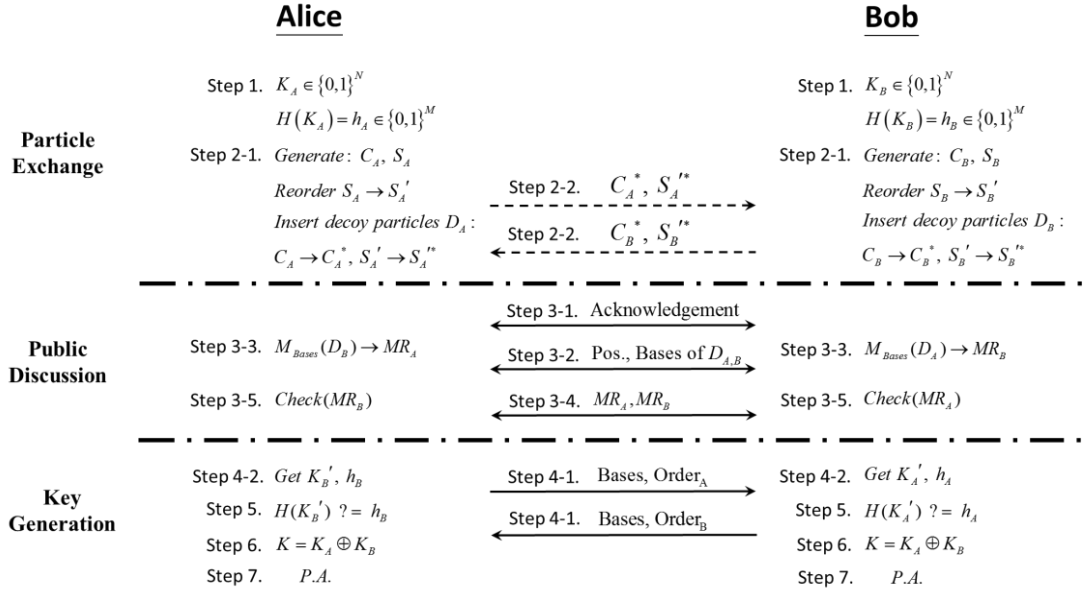

Figure 2. Proposed QKA with fairness

**Particle Exchange Stage:**

**Step 1**  Alice and Bob separately determine an *n*-bit string $K_A$ and $K_B$ as their private key. They use the hash function $H$ to generate *m*-bit hash values

$h_A = H(K_A)$ and $h_B = H(K_B)$, respectively.

**Step 2** Alice generates two sequences of single particles $C_A$ and $S_A$ from $h_A$ and $K_A$ based on the pre-agreed rules respectively. Then, she randomly reorders the particles of $S_A$ (denoted as $S_A'$). Note that the original order of the sequence $S_A'$ is known to nobody but Alice. Besides, Alice prepares a sufficient number of decoy particles randomly in one of four polarization states $\{|0\rangle, |1\rangle, |+\rangle, |-\rangle\}$. She randomly inserts these decoy particles into the sequences $C_A$ and $S_A'$ (denote the new sequences as $C_A^*$ and $S_A'^*$). After that, Alice sends the sequences $C_A^*$ and $S_A'^*$ to Bob via a quantum channel. Following the same process, Bob sends Alice two sequences $C_B^*$ and $S_B'^*$ as well.

**Public Discussion Stage:**

**Step 3** After receiving the sequences, Alice and Bob acknowledge each other via an authenticated classical channel. Then, they publish the positions and the measurement bases of the decoy particles. Based on the published information, the receiver measures the decoy particles and subsequently returns the measurement results back to the original sender. Now, Alice and Bob check whether the received measurement results are matched to the initial states they prepared. If the error rate is higher than a pre-defined threshold, the protocol will be aborted. Otherwise, Alice and Bob can confirm that the channel is free from the eavesdropping attacks.

**Key Generation Stage:**

**Step 4** Alice announces the bases of the remaining particles (i.e., $C_A$ and $S_A'$) and the original order of $S_A'$ to Bob. Bob recovers the sequence $S_A$ from $S_A'$ and measures $S_A$ and $C_A$ based on Alice's announcement. Therefore, Bob can derive Alice's $K_A'$ and $h_A$. Follow the same process, Alice can also obtain $K_B'$ and $h_B$.

**Step 5** Alice and Bob respectively obtain $h_B'$ from $H(K_B')$ and $h_A'$ from $H(K_A')$ and then compare the results to $h_B$ and $h_A$, respectively. In this regard, if there is a participant who finds that his/her comparison result is negative, he/she can deduce that the other one performs some illegitimate operations to affect the negotiated key (sending a wrong permutation, for instance). Otherwise, he/she can confirm the derived key is correct and reliable.

**Step 6** Alice and Bob can construct a shared raw key as $K' = K_A \oplus K_B$.

**Step 7** Finally, Alice and Bob obtain the final secret key $K = Privacy.Amplification.(K')$.

## 3.3 Security analyses and fairness analyses

Here, we show that the proposed QKA is secure against both the outsider eavesdropping attack and insider attack. In addition, we discuss how the proposed scheme satisfies the requirements of the strong fairness property.

### 3.3.1 Security against outsider attack

In this part, we use several well-known attacks (measure and resend attack, correlation-elicitation attack) as examples to show that the proposed QKA is secure against the outsider eavesdropping attack. Moreover, we prove that the proposed protocol is free from information leakage.

**a. Intercept and Resend Attack**

Suppose that there is an outside eavesdropper, Eve, who possesses powerful quantum capability and intends to acquire the shared secret key between Alice and Bob. In this regard, Eve has to obtain Alice and Bob's private key. More precisely, Eve intercepts the sequences sent from Alice and Bob in **Step 2** and performs measurements on the sequences to retrieve the private keys $K_A$ and $K_B$. Then, she sends the fake quantum sequences which have the same values with Eve's measurement results instead in hope to pass the eavesdropping check of the **Public Discussion Stage**. However, Eve has no idea about the positions of the decoy particles as well as the measurement bases of the sequences. For each decoy particle in one of the four polarization states $\{|0\rangle, |1\rangle, |+\rangle, |-\rangle\}$, there is a probability of $\frac{3}{4}$ that a participant can get the correct measurement result even if Eve accidentally measures it (in the Z-basis or X-basis). Therefore, Eve will pass the public discussion with a significant probability in **Step 3**. However, if there are numerous decoy particles (denote the number as $l$) suffering from Eve's attack, the probability that Eve can pass the check has become $\left(\frac{3}{4}\right)^l$. In other words, Eve will be detected with a probability closed to 1 if the number $l$ is large enough.

**b. Correlation-Elicitation Attack**

In case of Correlation-Elicitation (CE) attack, the eavesdropper Eve may try to intercept $S_A^{'*}$ and $S_B^{'*}$ as the controlled photons and subsequently prepares some auxiliary photons as the target photons, to obtain the useful information of Alice and Bob's private key. For example, Eve entangles her auxiliary photons with $S_A^{'*}$ by performing the controlled-NOT (CNOT) operation on each two photons, i.e. use one photon of $S_A'^*$ to be the control bit and use a qubit in the state of $|0\rangle$ to be the target bit. If the control bit is in Z-basis, we cannot detect this attack. Because in this case the control bit cannot be changed by the CNOT operation. And if the control bit is in X-basis, we have the probability of $\frac{1}{2}$ to detect the eavesdropping. Because in this case, the control bit has been entangled with the target bit. Obviously, there is the probability of $\frac{1}{2}$ to get the same measurement result of the control bit as before and the probability of $\frac{1}{2}$ to get a different measurement result. Overall, for one decoy photon, the detection rate is $\frac{1}{4}$. Hence, the probability that Eve can pass the check is $\left(\frac{3}{4}\right)^l$, which means that Eve will be detected with a probability closed to 1 if the number $l$ is large enough.

**c. Information Leakage Analysis**

The security of the final shared key is very important for the participants, and information leakage is a kind of passive attack that enables Eve to extract the secret key from the measurement results. In the following, we demonstrate that the proposed QKA can prevent the information leakage problem.

In the proposed protocol, if Eve wants to eavesdrop any useful information, he must intercepts the $S_A'^*$ and $S_B'^*$ in **Step 2**. Similar as the analysis of **Intercept and Resend Attack**, for one bit, the probability for Eve to obtain the correct measurement result is $\frac{3}{4}$. Hence, the measurement result of one qubit contains $1-\sum p_i \log_2 p_i = 1 - \frac{3}{4}\log_2\frac{3}{4} - \frac{1}{4}\log_2\frac{1}{4} \approx 0.1887$ bit of information. That means, according to the measurement results, Eve can obtain about 18.87% of Alice and Bob's private key. However, the analysis of **Intercept and Resend Attack** shows that this attack can be detected in the **Step 3**. Moreover, in the **Step 7**, the privacy amplification can ensure there is no information leakage in the proposed protocol. For example, if we assume the privacy amplification remains 80% of the raw key $K'$ to be the final secret key $K$. In other words, the reduced part is 20% which larger than 18.87%. Obviously, the proposed quantum key agreement protocol is free from information leakage.

### 3.3.2 Security against insider attack

This study points out that most of the QKAs suffer from the key manipulation problem as discussed in Section 2. Now we propose several possible scenarios to prove that our QKA can resist such an insider attack. In other words, the proposed protocol can achieve the strong fairness property.

First, suppose that there is a legitimate but malicious participant, Bob, who wants to control the shared key in his favor. After receiving the positions and the bases of the decoy particles from Alice in Step 3, Bob can draw the sequences $S_A'$ and $C_A$ from $S_A'^*$ and $C_A^*$ respectively. However, Bob has no idea about the original order of $S_A'$ and the measurement bases. He cannot efficiently extract Alice's private key $K_A$ and get any useful information from $S_A'$ and $C_A$ at this step. Thus, what Bob can do is only to follow the protocol and complete the public discussion honestly.

Second, consider that Bob tries to learn a little part of $K_A$. In the extreme case, merely one bit. In Step 3, after extracting the sequences $S_A'$ and $C_A$ according to Alice's announcement, Bob directly measures a particle of $S_A'$ in the Z-basis or X-basis at random). In this case, there is a probability of $\frac{3}{4}$ that he can derive a correct bit of $K_A$. However, the particles of the sequence $S_A'$ have been randomly reordered by Alice. Even if Bob gets a correct key bit, he still does not know which position that the bit belongs in $K_A$. Therefore, Bob cannot accurately learn any bit of $K_A$.

Third, after Alice and Bob complete the public discussion in Step 3, they can confirm the channel is free from the eavesdropping attacks. It indicates that the eavesdropper is eliminated in the protocol. Under this assumption, assume that Alice first announces the original order and corresponding measurement bases to Bob. In this case, Bob can immediately produce $K_A$ and further derive the shared secret key (by $K_A \oplus K_B$). Once Bob does not prefer the shared key, he will try to modify his own key $K_B$ by announcing a fake order to Alice. However, Alice will definitely get incorrect $K_B'$ and the error will be detected in Step 5. As a consequence, Alice can reasonably deduce that Bob reports a wrong permutation to change the shared key. Bob's illegitimate operation will be caught by Alice.

### 3.4 Strong Fairness

Section 3.1 described how to achieve the strong fairness property in a QKA. In the case of our proposed scheme, participants cannot obtain any useful information (other participants' private keys or hash values) to derive the final shared secret key until the public discussion finished. Hence, according to the proposed solution model, if a malicious participant tries to deliberately abort the protocol to manipulate the final key, then with a high probability he/she would be detected by the other participant.

## 4 Conclusion

This paper has pointed out that, according to the ISO/IEC DIS 11770-3, the existing definition of 'fairness' in QKAs is not rigorous and proposed the definition of the weak fairness property and the strong fairness property. Besides, most of the existing QKAs suffer from the key manipulation problem, which further causes these protocols unable to satisfy the strong fairness property. In this regard, we have proposed a model which provides a way of designing a QKA with the strong fairness property. Furthermore, we have presented a two-party QKA based on the solution model. Security analysis shows that the proposed QKA is secure against both the outsider and insider attacks, and also can avoid the key manipulation problem. Moreover, the same strategy described in Section 2.2 can also be applied to the construction of fair probabilistic quantum key distribution [43].

Finally, it is noteworthy to mention an issue which should be considered in the proposed two-party QKA. In Step 4 of the protocol, even the misconduct of the malicious participant, Bob, will certainly be detected by Alice in the protocol, Alice has no way to prove Bob's misconduct to a third person. As a consequence, this dispute cannot be solved adequately. Hence, how to solve this problem will be an interesting open problem.

## Acknowledgement

We would like to thank the Ministry of Science and Technology of the Republic of China, Taiwan for partially supporting this research in finance under the Contract No. MOST 105-2221-E-006 -162 -MY2.